\documentclass[aps,prd,nofootinbib,
twocolumn,
superscriptaddress,floatfix]{revtex4}
 
\usepackage{amsmath,amssymb}
\usepackage[dvips, pdftex]{graphicx}
\usepackage[tight,footnotesize]{subfigure}
\usepackage[usenames]{color}
\usepackage[dvipsnames]{xcolor}
\usepackage{float}
\usepackage{wrapfig}
\usepackage{hyperref}
\usepackage{mathtools}
\usepackage[utf8]{inputenc}
\usepackage{graphicx}
\usepackage{physics}

\usepackage{bm}
\usepackage{braket}
\usepackage{listings}
\usepackage{cases}
\usepackage{comment}
\usepackage{soul}
\usepackage{cancel}
\usepackage{cases}
\usepackage{xspace}
\usepackage{acronym}
\usepackage{siunitx}

\usepackage{xcolor}

\newcommand{\rhofl}{\rho_{\rm fl}}

\newcommand{\bea}{\begin{eqnarray}}
\newcommand{\eea}{\end{eqnarray}}

\acrodef{CMB}{cosmic microwave background}
\acrodef{PBH}{primordial black hole}
\acrodef{PDF}{probability density function}
\acrodef{EoM}{equation of motion}
\acrodef{GW}{gravitational wave}
\acrodef{RD}{radiation-dominated}
\acrodef{DM}{dark matter}
\acrodef{PTA}{pulsar timing array}
\acrodef{SIGW}{scalar induced gravitational wave}
\acrodef{SGWB}{stochastic gravitational wave background}
\acrodef{BSSN}{Baumgarte--Shapiro--Shibata--Nakamura}
\acrodef{LISA}{Laser Interferometer Space Antenna}
\acrodef{NANOGrav}{North American Nanohertz Observatory for Gravitational Waves}
\acrodef{DECIGO}{Decihertz Gravitational Wave Observatory}
\acrodef{FLRW}{Friedman–Lemaitre–Robertson–Walker}

\newcommand{\hD}{\hat{\Delta}}

\newcommand{\be}{\begin{equation}}
\newcommand{\beqn}{\begin{eqnarray}}
\newcommand{\eeq}{\end{equation}}
\newcommand{\eeqn}{\end{eqnarray}}

\begin{document}
\title{
Primordial Black Holes from Primordial Voids
}

\author{Cristian Joana}
\email{cristian.joana@itp.ac.cn}
\affiliation{International Centre for Theoretical Physics Asia-Pacific,
University of Chinese Academy of Sciences, 100190 Beijing, China}
\affiliation{Institute of Theoretical Physics,
Chinese Academy of Sciences, Beijing 100190, China}

\author{Zi-Yan Yuwen}
\email{ziyan.yuwen@apctp.org}
\affiliation{Asia Pacific Center for Theoretical Physics (APCTP), Pohang 37673, Korea}
\affiliation{Institute of Theoretical Physics,
Chinese Academy of Sciences, Beijing 100190, China}

\date{\today}

\begin{abstract}
Primordial black holes (PBHs) are a compelling dark matter candidate and a unique probe of small-scale cosmological fluctuations. Their formation is usually attributed to large positive curvature perturbations, which collapse upon Hubble re-entry during radiation domination. In this work we investigate instead the role of \emph{negative curvature perturbations}, corresponding to the growth of primordial void (PV) like regions. Using numerical relativity simulations, we show that sufficiently deep PV can undergo a nonlinear rebounce at the center, generating an effective overdensity that eventually collapses into a PBH. We determine the critical threshold for this process for a variety of equations of state, and demonstrate that the resulting black holes masses obey a scaling relation analogous to the standard overdensity case. These results establish primordial voids as a novel channel for PBH formation and highlight their potential impact on PBH abundances and cosmological signatures.
\end{abstract}

\maketitle

\section{Introduction} 

Primordial black holes (PBHs)~\cite{Hawking:1971ei,Carr:1974nx} are among the most promising candidates to explain the nature of dark matter, while also providing a unique probe of physics in the early Universe. Unlike astrophysical black holes, PBHs arise from non-stellar processes operating shortly after inflation, from processes that include the collapse of large adiabatic and iso-curvature perturbations~~\cite{Shibata:1999zs,Passaglia:2021jla,Yoo:2021fxs}, cosmic string loops~\cite{Jenkins:2020ctp,Helfer:2018qgv}, or bubble collisions~\cite{Jung:2021mku} (for reviews see \cite{LISACosmologyWorkingGroup:2023njw,Escriva:2022duf,Carr:2020xqk}). Among all possibilities, the most widely studied and theoretically well-motivated scenario involves the collapse of curvature fluctuations during the radiation-dominated era~\cite{Polnarev:2006aa,Escriva:2019phb,Escriva:2020tak}, other pressure(less) fluids~\cite{Mack:2006gz,Harada:2016mhb}, and scalar fields~\cite{deJong:2021bbo,deJong:2023gsx,Cheng:2025eas,Milligan:2025zbu,Padilla:2025bkv}. These perturbations are typically seeded by quantum fluctuations during inflation, stretched to super-Hubble scales, which later re-enter the Hubble volume. 

In this standard picture, attention has centered on positive curvature perturbations, corresponding to overdense regions of spacetime. For such profiles, the local density contrast grows non-linearly upon Hubble re-entry, and collapse occurs if a critical threshold is exceeded \cite{Carr:1975qj,Musco:2004ak,Musco:2012au,Escriva:2019phb,Escriva:2020tak}.
This specific mechanism underlies the bulk of PBH formation models, and 
the potential observational consequences of PBHs, such as their role as dark matter, seeds for supermassive black holes, or sources for gravitational waves~\cite{Sasaki:2018dmp,Carr:2020xqk,LISACosmologyWorkingGroup:2023njw}, are therefore predominantly assessed within this framework.

\begin{figure}[t]
    \centering
    \includegraphics[width=0.95\linewidth]{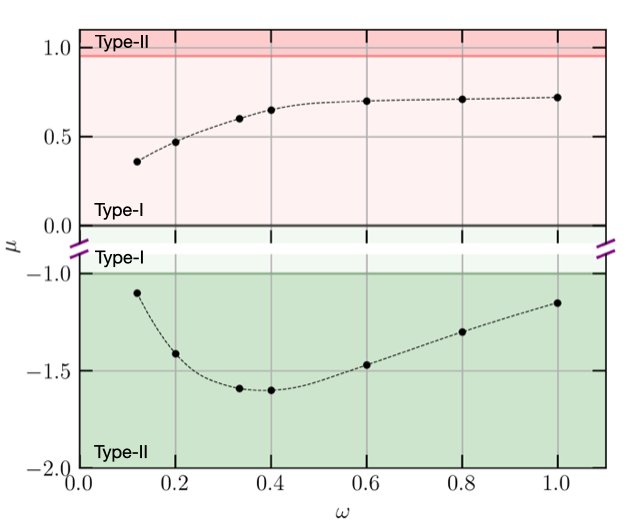}
    \caption{Amplitude thresholds for type-A PBH formation across different equations of state $\omega$. Black dots denote results from numerical relativity simulations, connected by a solid black line (quadratic interpolation) representing the separatrix. The light and dark red (green) shaded regions correspond to type-I and type-II classifications for positive (negative) curvature fluctuations, respectively.}
    \label{fig:thresholds}
\end{figure}

By contrast, the role of \emph{negative curvature perturbations} has received little attention. These fluctuations correspond to underdensities or void-like regions in the primordial plasma, which we refer to as Primoridal Voids (PVs). At first glance, they appear unlikely to collapse, since their geometry counteracts gravitational attraction. However, once the nonlinear dynamics of Hubble re-entry are taken into account, deep negative curvature perturbations may evolve in a qualitatively different way. Rather than simply dispersing, such void-like profiles can generate overdense shell-structures that, in turn, might drive gravitational collapse and produce PBHs. This provides a new and unexplored channel for PBH formation, complementing the standard overdensity case.

The study of this mechanism requires going beyond linear perturbation theory. The general relativistic evolution of such void-like perturbations involves strong nonlinearities both in the metric and in the fluid sector, which cannot be captured analytically in a closed form. Numerical relativity offers the necessary framework to follow these dynamics from Hubble entry through to possible collapse, while analytical arguments can guide the identification of thresholds and scaling laws. Together, these tools enable a systematic exploration of PBH formation from negative curvature perturbations.

In this work we carry out a dedicated study on PBH formations from rebounces of PVs. We identify the critical amplitude of negative curvature fluctuations required for collapse, demonstrate that the resulting PBHs masses obey a scaling relation akin to the overdensity case, and map the competing outcomes of dispersal, shock formation, and black-hole collapse. Extending on \cite{Germani:2025hcu}, these results open a new avenue for PBH phenomenology and highlight the need to include PVs when assessing PBH abundances and their observational consequences.

The paper is organized as follows. In Section~\ref{sec:curvature}, we review the theoretical framework of PBH formation from curvature fluctuations. Section~\ref{sec:GRHydro} details our numerical methodology based on general relativistic hydrodynamics. Our main results on collapse dynamics, critical thresholds, and mass scaling are presented in Section~\ref{sec:results}. The conclusions are left for Section~\ref{sec:conclusions}. Throughout this work, we use the ``mostly-plus'' metric signature $(-,+,+,+)$ and assume geometrical units where $G = c = 1$.

\section{Primordial black holes from curvature fluctuations}\label{sec:curvature}

Large curvature perturbations in the early Universe can arise from inflationary dynamics beyond the simplest slow-roll picture. In particular, non-attractor phases provide a natural way to amplify the curvature perturbation $\zeta$ on super-Hubble scales. Examples include ultra-slow roll inflation~\cite{Germani:2017bcs,Figueroa:2020jkf,Pattison:2021oen}, constant-roll inflation~\cite{Motohashi:2019rhu,Motohashi:2017aob,Motohashi:2017vdc,Tomberg:2023kli,Inui:2024sce}, or spectator field scenarios such as curvaton dynamics~\cite{Pi:2021dft}. In these cases, the usual conservation of $\zeta$ outside the Hubble volume is broken, allowing its amplitude to grow and potentially reach the large values required for PBH formation.  
A useful organizing principle for these scenarios is provided by the so-called \emph{logarithmic duality}~\cite{Atal:2019cdz,Pi:2022ysn,Inui:2024fgk,Inui:2024sce}, which relates the non-Gaussian curvature perturbation to an underlying Gaussian field and connects various non-attractor mechanisms. Explicitly, the curvature takes the form  
\begin{equation} \label{eq:profile_NG}
\zeta_{\rm log}(\vec x) \;=\; -\frac{1}{\gamma_{\rm NG}} \, \log\!\left( 1 - \gamma_{\rm NG}\,\zeta_G(\vec x) \right) \, ,
\end{equation}
where $\zeta_G$ denotes the Gaussian curvature perturbation and $\gamma_{\rm NG}$ parametrizes the strength and sign of non-Gaussian effects. Besides modifying the mean curvature profile, this relation also alters the probability distribution function. Models with $\gamma_{\rm NG} > 0$ introduce a positive skewness that enhances the abundance of overdensities, whereas models with $\gamma_{\rm NG} < 0$ enhance the population of voids.

For simplicity, however, we adopt a set of assumptions that notably reduce the complexity of the problem.  First, we restrict ourselves to the Gaussian case ($\gamma_{\rm NG} =0$), ignoring non-Gaussian modification to the shape profile. Second, we assume a narrow-peak (monochromatic) power spectrum, with amplitude $A_\zeta$ peaked at $k_\star$,
\begin{equation}
\mathcal{P}_\zeta(k) = A_\zeta \delta(k - k_\star)~,
\end{equation}
which reduces scale dependence of the system to $k_\star$. Finally,
we focus on the rare large peaks (maxima or minima) where the spherical symmetry approximation is justified. Under these assumptions, the mean radial profile of a curvature perturbation is given by a sinc-function~\cite{Bardeen:1985tr,Yoo:2018kvb,Yoo:2020dkz},
\begin{equation}\label{eq:sincfunction}
\zeta(r) \;=\; \mu \, \mathrm{sinc}\!\left(k_\star r\right) \, ,
\end{equation}
where $\mu$ is the central amplitude. The study of other shape profiles—such as those linked to a broad-peak power spectrum or the effects of non-Gaussianities—is beyond the scope of this paper and is left for future work. Here, we focus on the profile given by Eq.~(\ref{eq:sincfunction}) and use it to generate the initial conditions for the numerical relativity simulations presented in Section~\ref{sec:results}.

\subsection{Geometric classifications of fluctuations and dynamical collapse}

\begin{figure*}[t]
    \centering
    \includegraphics[width=0.95\linewidth]{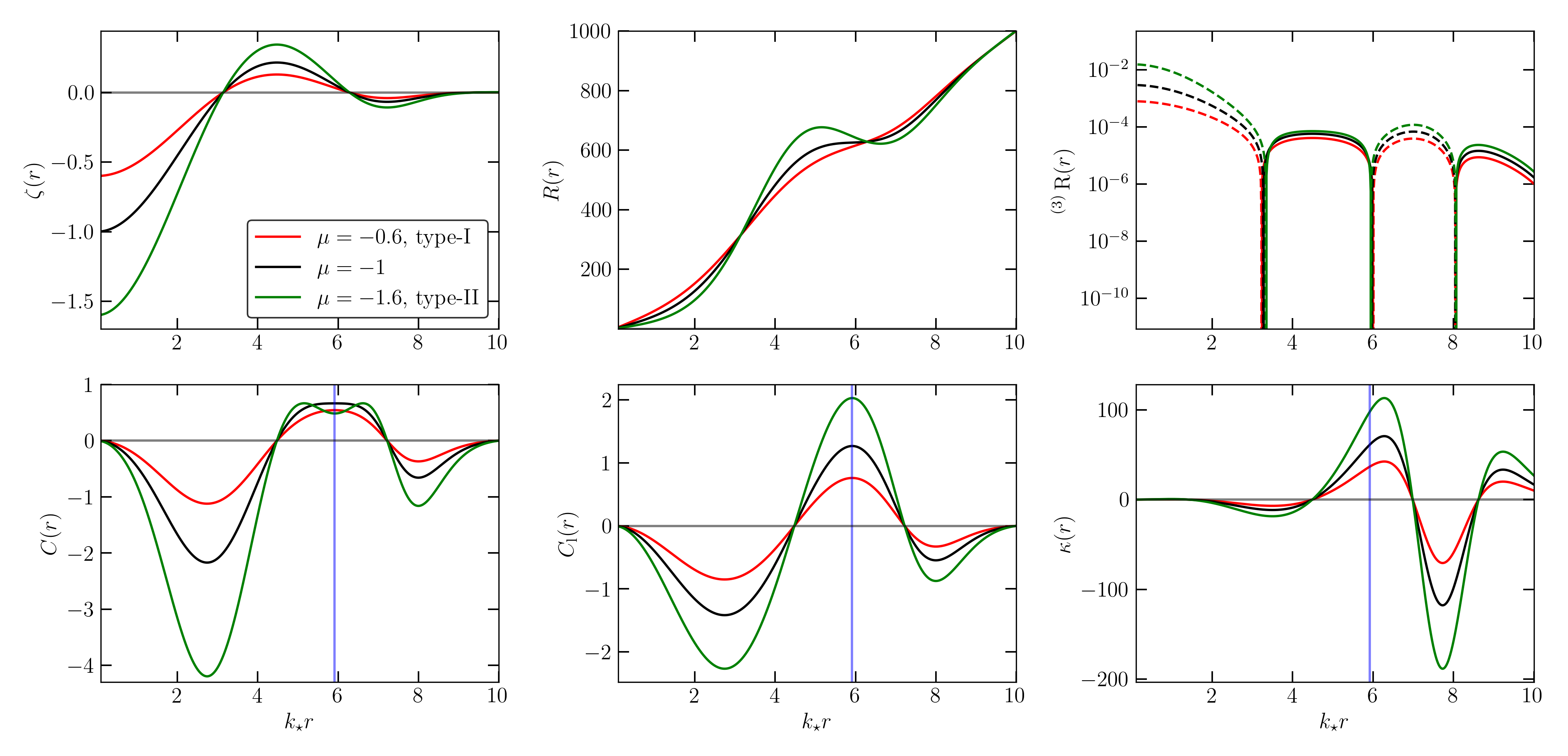}
    \caption{Initial configuration for three representative negative curvature fluctuations with amplitudes $\mu = -0.6$ (red), $-1$ (black), and $-1.6$ (green). \textbf{Top panels:} the initial curvature profile $\zeta(r)$, the areal radius ${R}(r)$, and the three-Ricci scalar ${}^{(3)}R(r)$. In the latter panel, positive values are represented with solid lines, whereas negative values with dashed line. \textbf{Bottom panels:} the corresponding non-linear compaction function $\mathcal{C}(r)$, its linear component, and the compaction's curvature $\kappa(r)$, computed assuming radiation domination ($\omega=1/3$). The chosen amplitudes exemplify the type-I (red), type-I/II transitional (black), and type-II (green) regimes. The vertical blue line in the bottom panels indicate the location of the maximum of $C_\ell$.}
    \label{fig:geometry}
\end{figure*}

An important quantity for characterizing the geometry of a curvature fluctuation is the areal radius, defined as
\begin{equation}
R(r) = \mathfrak{a}(t) e^{\zeta(r)} r ,
\end{equation}
where $r$ is the comoving radius and $\mathfrak{a}(t)$ is the cosmological scale factor. Using the notation from Refs.~\cite{Kopp:2010sh,Uehara:2024yyp,Harada:2024trx}, we can classify super-Hubble curvature fluctuations as either type I or type II. In type I, $R(r)$ is monotonic, such that $\mathrm{d}R/\mathrm{d}r>0$ everywhere, whereas in the type II case, the areal radius develops a non-monotonic region where $\mathrm{d}R/\mathrm{d}r<0$. This non-monotonicity leads to a warped geometry, developing a "belly" and "neck" structure, where the angular distance between diametrically opposite points inside the "neck" becomes shorter than the radial distance between them.
For the sinc profile with $\mu > 0$, the transition between type I and type II occurs at $\mu \gtrsim 0.95$. Notably, negative amplitude fluctuations also admit type II configurations, which occur for $\mu < -1$ (see Fig.~\ref{fig:geometry} for representative examples).

It is important to stress that the classification into type I and type II refers to the \emph{geometry of the super-Hubble fluctuation}, while the recently introduced classification into type-A and type-B instead concerns a \emph{dynamical classification of the collapse process} of a PBH. Type-A PBHs form after Hubble re-entry, when the central overdensity grows sufficiently to overcome pressure support and collapse, eventually creating an apparent horizon. Type-B PBHs, in contrast, form through a "separate universe" mechanism. Here, a sufficiently strong curvature fluctuation creates a causally disconnected region from the background expansion, which develops a trapped surface before being incorporated into the Hubble volume of the expanding background (see \cite{Shimada:2024eec,Uehara:2024yyp} for recent studies). In this sense, type A/B is a dynamical classification of the collapse process, whereas type I/II is a geometric classification of the curvature profile at super-Hubble scales.

As recently proposed in ~\cite{Germani:2025hcu}, an additional relevant classification concerns the sign of the spatial curvature at the fluctuation's core, providing a complementary geometric characterization than the Type I/II classification. We can use the three-dimensional Ricci scalar to classify the fluctuation's core geometry as open, flat, or closed. At super-Hubble scales, the three-Ricci scalar reads
\begin{equation}
{}^{(3)} R = - e^{-2\zeta(r)} \left( 4 \zeta''(r) + \frac{8}{r} \zeta'(r) + 2 \left(\zeta'(r)\right)^2 \right).
\end{equation}
This classification is relevant for PBH formation, because a closed core would facilitate the collapse, leading to a lower critical amplitude threshold, whereas an open core would oppose it.
In this light, positive and negative values of $\mu$ in Eq.(\ref{eq:sincfunction}) correspond to closed and open core-types, respectively, and therefore a higher collapse threshold is expected for the PVs sector.

\subsection{Trapped Surfaces and Misner--Sharp Mass in an Expanding Universe} \label{subsec:trappedregions}

The local geometry of a spacetime hypersurface can be characterized by the null expansions of the outgoing and ingoing radial null congruences,  $\Theta_{+}$  and $\Theta_{-}$, respectively. 

For instance, in the pure expanding FLRW universe, the null expansions are defined as
\begin{equation}
    \Theta_{\pm} = \frac{2}{R}\left( H R \pm 1 \right)~,
\end{equation} 
and therefore, the structure of null congruences reads
\begin{align}
    &\Theta_+ > 0, \ \Theta_- < 0 \quad && \text{for} \quad HR < 1, \\
    &\Theta_+ > 0, \ \Theta_- = 0 \quad && \text{for} \quad HR = 1, \\
    &\Theta_+ > 0, \ \Theta_- > 0 \quad && \text{for} \quad HR > 1.
\end{align}
The first case corresponds to an untrapped region.
The second case, at $\Theta_{-}=0$, defines the marginally outer trapped surface, which coincides with the Hubble radius. 
For $HR>1$, the $\Theta_{-}$ become positive, corresponding to a past-trapped region.

A covariant characterization of the gravitational energy contained within a volume is provided by the {Misner--Sharp mass} \cite{1964PhRv..136..571M,1966PhRv..141.1232M}, defined as
\begin{equation}
    M_{\mathrm{MS}} = \frac{R}{2}\left( 1 - g^{\mu\nu} \partial_\mu R \, \partial_\nu R \right).
\end{equation}

The Misner--Sharp mass satisfies the geometric relation
\begin{equation}
    1 - \frac{2M_{\mathrm{MS}}}{R} = g^{\mu\nu} \partial_\mu R \, \partial_\nu R,
\end{equation}
which vanishes at marginally trapped surfaces. 
Thus, the condition
\begin{equation} \label{eq:AHcondition}
    \frac{2M_{\mathrm{MS}}}{R} = 1
\end{equation}
defines the existence of an apparent horizon (AH), corresponding to the null expansion condition $\Theta_+ \, \Theta_- = 0$. 
Thus, the condition (\ref{eq:AHcondition}) is satisfied both at the Hubble distance of the cosmological background for the  $\Theta_{-}=0$ case  (a marginally past-trapped surface), and at the local AH of a forming PBH corresponding to $\Theta_{+}=0$ (a marginally future-trapped surface).\\

After a PBH is formed, the causal structure of null congruences changes towards to
\begin{align}
    &\Theta_+ < 0, \ \Theta_- < 0 \quad && \text{for} \quad R < R_{\text{PBH}}, \\
    &\Theta_+ = 0, \ \Theta_- < 0 \quad && \text{for} \quad R = R_{\text{PBH}}, \\
    &\Theta_+ > 0, \ \Theta_- < 0 \quad && \text{for} \quad R_{\text{PBH}} < R < R_H, \\
    &\Theta_+ > 0, \ \Theta_- =0 \quad && \text{for} \quad R = R_H, \\
    &\Theta_+ > 0, \ \Theta_- > 0 \quad && \text{for} \quad R > R_H,
\end{align}
where $R_{\text{PBH}}$ denotes the AH of the PBH, and $R_H$ the Hubble radius.

\subsection{Compaction function and threshold criterion}

An alternative diagnostic for identifying the formation of a black hole is the compaction function~\cite{Shibata:1999zs}, which can be expressed as
\begin{equation}
C(r) = \frac{2[M_{\rm MS}(r) - M_{\rm bkg}(r)]}{R(r)},
\end{equation}
where $M_{\rm bkg}(r)$ is the FLRW background mass within the same radius. When $C(r) \gtrsim 1$, a trapped surface forms, signaling black hole formation.

An important application of the compaction function is that its peak value at superHubble scales can be used to predict whether positive amplitude fluctuation will collapse into a black hole upon Hubble re-entry. Following Refs.~\cite{Germani:2023ojx}, one can define the linear compaction function as 
\begin{equation}
C_\ell(r) = -2\phi(w)\, r \, \zeta'(r) \, , 
\quad 
\phi(w) \equiv \frac{3(1+w)}{5+3w} \, ,
\end{equation}
which depends only on the profile of $\zeta(r)$ and the equation-of-state parameter $w$. The full (non-linear) compaction function is then given by
\begin{equation}
C(r) = C_\ell  \left(1- \frac 1{4\phi(w)} C_\ell\right).
\end{equation}

It is useful to introduce the curvature of the compaction function,
\begin{equation}
\kappa(r) \equiv - r^2 \frac{\mathrm{d}^2 C_\ell}{\mathrm{d}r^2} ~,
\end{equation}
and to evaluate this quantity at the radius $r_m$ where $C_\ell(r)$ attains its maximum. Let us use the notation $C_m \equiv C_\ell(r_m)$ and $\kappa_m \equiv \kappa(r_m)$ for brevity.

It has been shown in \cite{Escriva:2020tak,Escriva:2019phb} that the critical threshold for primordial black hole formation follows from a non-linear relation between $C_m$ and $\kappa_m$. This yields an implicit criterion of the form
\begin{equation} \label{eq:prescription}
C_m > \delta_c(C_m, \kappa_m),
\end{equation}
where $\delta_c(C_m, \kappa_m)$ is a known non-linear relation (see \cite{Escriva:2025rja} for the type II case). This approach has been shown to reproduce the thresholds measured in full numerical-relativity simulations with high accuracy (within the 5\%) for the radiation domination case, i.e. $\omega = 1/3$.

\section{General Relativistic Hydrodynamics} \label{sec:GRHydro}

\begin{figure*}[t]
    \centering
    \includegraphics[width=1.\linewidth]{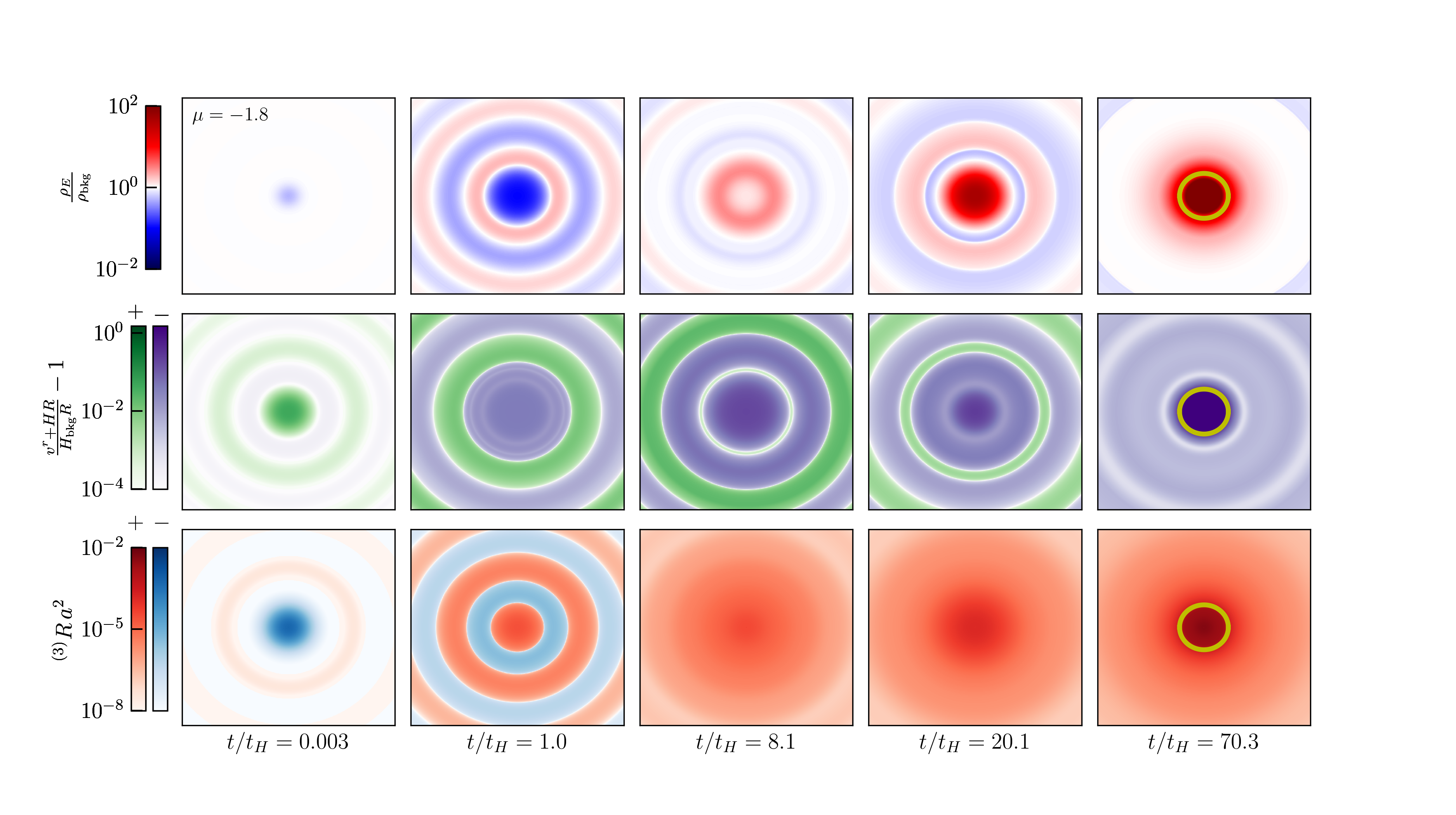}
    \includegraphics[width=1.\linewidth]{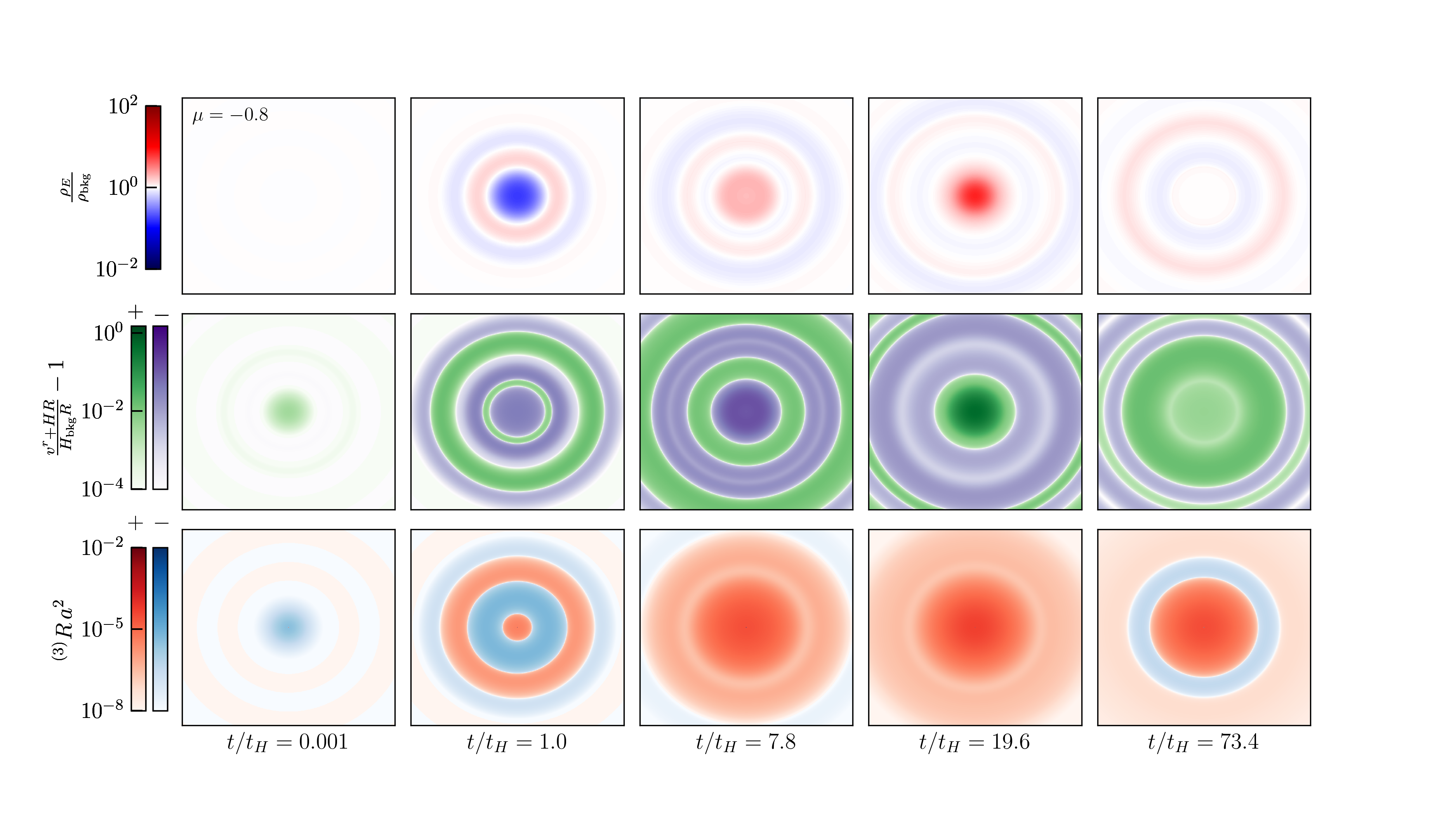}
    \caption{Evolution of the density contrast $\rho_E/\rho_{\rm bkg}$(upper rows), relative recession velocity with respect the background (middle rows) and normalized Ricci Scalar by a factor $\mathfrak{a}(t)^2$ (bottom rows) for negative curvature fluctuations re-entering the Hubble horizon during radiation domination $(\omega = 1/3)$, obtained from numerical relativity simulations. \textbf{Top panels:} initial fluctuation with amplitude $\mu = -1.8$ and $\omega = 1/3$, which forms a PV that rebounds into a central overdensity and collapses into a PBH. \textbf{Bottom panels:} initial fluctuation with amplitude $\mu = -0.8$ and $\omega = 1/3$,  which also forms a PV but whose rebound dissipates into sound waves. Underdensities are represented by a blue or purple colormaps, while overdensities by red or green colormaps. The yellow circle denotes the formation of an apparent horizon.}
    \label{fig:drho}
\end{figure*}

We now describe the general relativistic hydrodynamics methods used to evolve Hubble re-entry of negative curvature fluctuations, the formation of PVs and potential collapse into PBHs.

We describe the cosmological system using the Baumgarte-Shapiro-Shibata-Nakamura (BSSN) formalism~\cite{Baumgarte_1998,PhysRevD.52.5428}, reduced to spherical symmetry~\cite{Alcubierre:2011pkc}. Thus, the line element reads
\begin{equation}
ds^2 = -\alpha^2 dt^2 + e^{4\chi} \left[ a (dr + \beta^r dt)^2 + r^2 b\, d\Omega^2 \right],
\end{equation}
where $\Omega$ is the solid angle, $\alpha$ is the lapse function, $\beta^r$ is the radial component of the shift vector. The spatial metric is conformally decomposed $\gamma_{ij} = e^{4\chi} \tilde\gamma_{ij} $, where $\chi$ is the conformal factor and the metric components are defined by $a(r,t) \equiv \tilde{\gamma}_{rr}$ and $b(r,t) \equiv \tilde{\gamma}_{\theta\theta}/r^2$.

The extrinsic curvature $K_{ij}$ is decomposed into its trace, $K$, and its traceless part, $\tilde{A}_{ij}$. In this formulation, the mixed components $A_a \equiv \tilde{A}^r_r$ and $A_b \equiv \tilde{A}^\theta_\theta$ are used, which are related by the traceless condition $A_a + 2 A_b = 0$. The BSSN formalism also introduces the contracted Christoffel symbols of the conformal metric, $\hat{\Delta}^i \equiv -\partial_j \tilde{\gamma}^{ij}$, as independent variables, used for numerical stability in the simulations. 

We choose the matter sector to be described by a perfect fluid with stress-energy tensor
\begin{equation}
T_{\mu\nu} = (\rhofl + p) u_\mu u_\nu + p g_{\mu\nu},
\end{equation}
where $\rhofl$ and $p$ are the fluid's energy density and pressure, respectively, and $u_\mu$ is the four-velocity. We employ a barotropic equation of state $p = \omega \rhofl$, considering values of $\omega$ spanning from $0 < \omega < 1$, including the radiation domination case with $\omega = 1/3$. For a comoving observer, the energy density splits into rest energy density $\rho_0$ and internal energy $\varepsilon$ as $\rhofl \equiv \rho_0 (1 + \varepsilon)$. The energy density measured by an Eulerian observer is given by
\begin{equation}
\rho_{\rm{E}} = (\rhofl + p) W^2 - p,
\end{equation}
where $W \equiv 1/\sqrt{1-v^2}$ is the Lorentz factor and $v^r$ is the radial component of the fluid's three-velocity.

The state of the fluid is described by the primitive variables: the rest mass density $\rho_0$, specific internal energy $\varepsilon$, pressure $p$, and three-velocity $v^r$. For numerical evolution, these are transformed into the conserved variables $D$, $S_r$, and $\mathcal{E}$, which are defined as
\begin{align}
 D &= \rho_0 W, \\
 S_r &= (\rhofl + p) W^2 v_r, \\
 \mathcal{E} &= (\rhofl + p) W^2 - p - D ~.
\end{align}
The system is evolved by coupling the BSSN equations for the gravitational variables $\{\chi, a, b, K, A_a, \hat{\Delta}^r, \alpha, \beta^r\}$ with the hydrodynamic equations for the conserved matter variables $\{D, S_r, \mathcal{E}\}$. The complete set of evolution equations is detailed in Ref.~\cite{10.1093/acprof:oso/9780199205677.001.0001,Alcubierre:2011pkc} and summarized in the Appendix. All simulations begin from constraint-satisfying initial conditions.   The methods for solving the initial data, evolution schemes and other numerical details can be found in the Appendix.

\section{Numerical results} \label{sec:results}

Our numerical simulations begin on an initial equal-time hypersurface containing a super-Hubble curvature fluctuation $\zeta(t,r)$ centered at the origin of the radial coordinate, embedded in a Friedmann–Lemaître–Robertson–Walker (FLRW) expanding background. The background scale factor is ${\mathfrak{a}(t)=\mathfrak{a}_0 (t/t_0)^{\sigma}}$, with Hubble rate $H_{\rm bkg}(t)=\sigma/t$, where ${\sigma = 2/[3(1+\omega)]}$ for a perfect fluid of equation-of-state parameter $\omega$. For convenience,  we set $t_0 = 1$ and $\mathfrak{a}_0 = \mathfrak{a}(t_0) = 1$.

The characteristic scale of the initial curvature fluctuation is fixed by  $k_\star^{-1} = 100\,H_{\rm bkg}^{-1}(t_0)$, ensuring that the perturbation is well outside the Hubble radius  at the start of the simulation. Thus, we define the Hubble-crossing time $t_H$ as the instant when  $\mathfrak{a}(t) H_{\rm bkg}(t) / k_\star = 1$.

\subsection{Dynamics of gravitational collapse}

\begin{figure*}[t]
    \centering
    \includegraphics[width=0.49\linewidth]{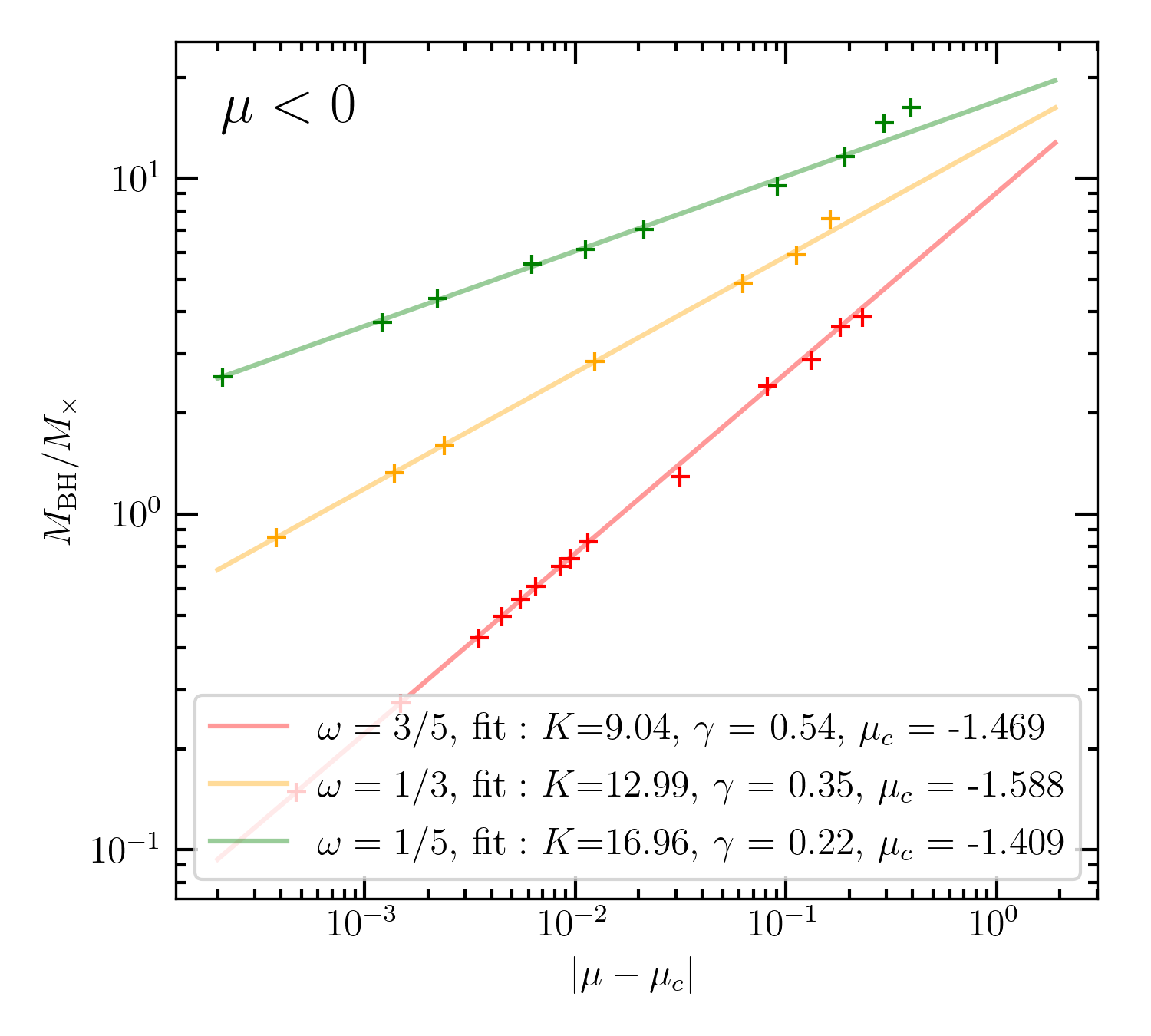}
    \includegraphics[width=0.49\linewidth]{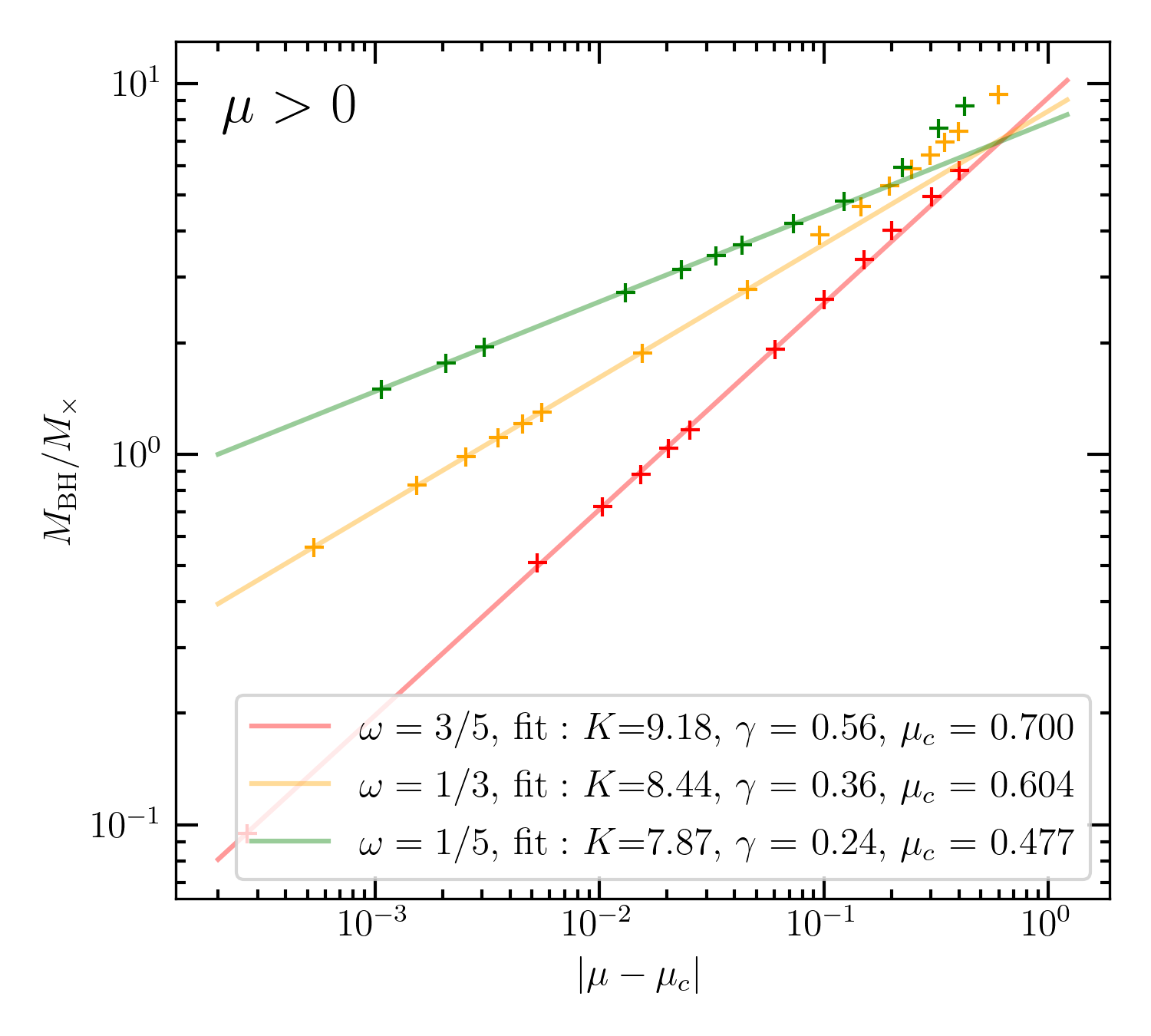}
    \caption{
    Mass scaling of PBH formation for negative (left panel) and positive (right panel) curvature fluctuations. 
    }
    \label{fig:bhmasses}
\end{figure*}

For completeness, we present results from both positive and negative $\zeta$ perturbations and compare them. 

For a positive amplitude $\mu$, the perturbation corresponds to an initial overdensity. While outside the Hubble horizon, the perturbation remains quasi-linear, i.e. following linear perturbation theory until $k_\star/\mathfrak{a} H_{\rm bkg} \sim 0.1$. Its subsequent evolution after horizon entry is characterized by rapid growth and strong self-gravity. The final outcome is then determined by the competition between this self-gravity and pressure gradients. For sufficiently large amplitude $\mu > \mu_c^{(+)}$, the gravity dominates and the region collapses to form a PBH, whereas when not, the pressure support is sufficient to prevent the collapse and the overdensity disperses into sound waves without forming a black hole.

In contrast, a negative amplitude $\mu$ corresponds to an initial underdensity. The central underdensity deepens progressively already on super-Hubble scales, where the negative curvature perturbation causes the inner region to expand faster than the outer shells, reaching its maximum depth at Hubble crossing—a stage we refer to as a PV. Owing to the oscillatory structure of the sinc profile, a PV is always accompanied by a surrounding overdense shell. During the subsequent sub-Hubble evolution, this shell contracts under its self-gravity and the background pressure, driving matter inwards and causing the central region to rebound. In this scenario, pressure initially drives the contraction, in direct contrast to the positive $\mu$ case where it always opposes collapse. For a sufficiently negative amplitude $\mu < \mu_c^{(-)}$, the rebound is sufficiently strong to collapse into a PBH; otherwise, it disperses, generating sound waves. This mechanism highlights the qualitative difference between PBH formation from overdensities and PVs. Figure~\ref{fig:drho} illustrates the dynamics of PVs in both collapsing and dissipating scenarios. The top rows show the energy density ratio $\rho_E/\rho_{\rm bkg}$ for times before Hubble crossing, at rebound, and during the final collapse or dissipation of the central feature. The middle rows show the relative recession velocity from the origin, i.e. $ V_{\rm rec} = v^r + HR $, where $H = -K/3 $ is the local expansion rate, with respect to the background recession $V_{\rm rec,\, bkg} = H_{\rm bkg} R$. Finally, in the bottom rows, we also show the evolution of the Ricci curvature, i.e.  $ ^{(3)}R\cdot \mathfrak{a}(t)^2$.

\subsection{Equation of state dependence and thresholds}

The formation dynamics exhibit a strong dependence on the fluid pressure, parameterized by the equation of state $\omega$. We have computed the critical thresholds for PBH formation (see Table~\ref{tab:critical_thresholds}) from both overdensities and PVs across a range of $\omega$, from the nearly pressureless limit ($\omega \simeq 0$) to the stiff fluid limit ($\omega=1$). Figure \ref{fig:thresholds} shows the numerical collapse thresholds $\mu_{\rm c}$ for representative $\omega$ values. As expected, the thresholds for PVs exceed those for overdensities, as gravitational collapse must overcome the expanding void geometry. Moreover, we find that while positive-$\mu$ perturbations can form PBHs directly from Type I fluctuations, negative-$\mu$ perturbations require Type II fluctuations. All PBHs produced in the vicinity after $\left|\mu\right| > \left|\mu_c^{(\pm)}\right|$ are of Type A.

Furthermore, the thresholds for positive $\mu$ increase monotonically with $\omega$, elucidating the role of pressure to prevent gravitational collapse. Instead, thresholds in the PV formation channel display a non-monotonic behavior, peaking near the radiation-dominated case $\omega \simeq 1/3$ and decreasing for both softer and stiffer equations of state. We interpret this as a reflection of competing effects of pressure gradients and sound speed in the collapse dynamics. For $\omega < 1/3$, the reduced fluid's pressure  facilitates the collapse once the rebounce is produced. Conversely, for $\omega > 1/3$, the increased sound speed associated with higher pressure, i.e. $c_s \approx~ \sqrt{\omega}$,  enables a faster compression rate, producing a more violent rebound that effectively facilitates PBH formation, even though pressure subsequently opposes collapse once the overdense core is generated.

Another example to illustrate the sound speed effect is on the near-dust limit ($\omega\to 0$). In this case, the overdense shell surrounding a PV barely contracts, but instead, the shell begins collapsing at nearly constant radius (i.e. the shell narrows in thickness rather than undergoing significant inward motion). Numerical limitations prevent definitive conclusions in this regime, but upon inspection, we see that the compaction function remains small at all times, suggesting that a PBH might never form. The fate of the thin shell formation would be presumably strongly conditioned by (aspherical) environmental effects which are not captured in our simulations.

After comparison of the numerical thresholds with those predicted by the analytical prescription (\ref{eq:prescription}), our numerical results confirm a good agreement for the positive-curvature fluctuations scenario in the radiation domination case, with deviations under 5\%. For negative-curvature fluctuations, however, we observe that the prescription predicts a lower threshold\footnote{
We use the prescription for type-II fluctuations from \ref{eq:prescription}, of which ${\delta_c = a\kappa_m^b} $ with ${a = 0.51962}$  and ${b = 0.26687}$.
} ($\mu_c^{\rm th}\sim -1.1$), inducing  approximately a 20\% error. This disagreement is not surprising as these analytical approaches do not capture the specific open-core geometry associated to PVs.

\begin{table}[b!]
    \centering
    \caption{Numerical critical thresholds for several equations of state. Precision is $\pm 0.01$.}
    \label{tab:critical_thresholds}
    \ \\[1mm]
    \begin{tabular}{c @{\hspace{2em}} r @{\hspace{2em}} r}
        \hline \hline \\[-1.8mm]
        $\omega$ & $\mu_c^{(+)}$ & $\mu_c^{(-)}$ \\[1mm]
        \hline  \hline \\[-2mm]
        0.12     & 0.36 & -1.10 \\
        0.20     & 0.47 & -1.41 \\
        $1/3$    & 0.60 & -1.59 \\
        0.40     & 0.65 & -1.60 \\
        0.60     & 0.70 & -1.47 \\
        0.80     & 0.71 & -1.30 \\
        1.00     & 0.72 & -1.15 \\[1mm]
        \hline
    \end{tabular}
\end{table}

\begin{table}[b]
\centering
\caption{Fitted parameters for different equations of state $\omega$ for $\mu>0$ and $\mu<0$.
\label{tab:fitvalues}
}
\ \\[1mm]
\begin{tabular}{@{\hspace{1em}}c@{\hspace{1em}}| @{\hspace{1em}} c @{\hspace{1em}}}
\hline \hline \\[-1.5mm]
$\omega$ & Fit parameters for $\mu > 0$ \\[1mm]
\hline  \hline \\[-2mm]
$3/5$ & $K = 9.18,\ \gamma = 0.56,\ \mu_c^{(+)} = 0.700$ \\[1mm]
$1/3$ & $K = 8.44,\ \gamma = 0.36,\ \mu_c^{(+)} = 0.604$ \\[1mm]
$1/5$ & $K = 7.87,\ \gamma = 0.24,\ \mu_c^{(+)} = 0.477$ \\[1mm]
\hline
\end{tabular}
\ \\[5mm]
\begin{tabular}{@{\hspace{1em}}c@{\hspace{1em}}| @{\hspace{1em}} c @{\hspace{1em}}}
\hline \hline \\[-1.8mm]
$\omega$ & Fit parameters for $\mu < 0$ \\[1mm]
\hline  \hline \\[-2mm]
$3/5$ & $K = 9.04,\ \ \gamma = 0.54,\ \mu_c^{(-)} = -1.469$ \\[1mm]
$1/3$ & $K = 12.99,\ \gamma = 0.35,\ \mu_c^{(-)} = -1.588$ \\[1mm]
$1/5$ & $K = 16.96,\ \gamma = 0.22,\ \mu_c^{(-)} = -1.409$ \\[1mm]
\hline
\end{tabular}
\end{table}

\subsection{Scaling of mass functions}

Finally, we compute the critical mass scaling for $\omega = 1/5, 1/3, 3/5$, corresponding to soft, radiation, and stiff equations of state, respectively. The numerical results are fitted to the standard critical scaling form~\cite{Choptuik:1992jv,Evans:1994pj,Musco:2012au},
\begin{equation}
   \frac{M_{\rm PBH}}{M_\times} = K \left|\mu - \mu_c\right|^\gamma ~,
\end{equation}
where $M_\times$ is the mass within the Hubble volume at the time of Hubble-crossing, $K$ is a fitted mass ratio, $\gamma$ is the critical exponent, and $\mu_c$ is the critical threshold. The best fit values are shown in Table~\ref{tab:fitvalues}. Notably, we confirm a very similar fitting values between positive and negative curvature fluctuations at a given $\omega$, particularly for the scaling index $\gamma$ (see also Figure~\ref{fig:bhmasses}). The fits are performed, as customary, for near-threshold amplitudes $\delta\mu \equiv \left|\mu - \mu_c\right| < 10^{-1}$, successfully recovering the well-known value of $\gamma \approx 0.36$ for radiation domination~\cite{Evans:1994pj}. We observe, however, that for larger amplitudes $\delta\mu > 0.1$, the resulting masses exceed the predictions of the critical scaling law, a deviation already seen in~\cite{Escriva:2022duf}.

\section{Conclusions and Discussions}
\label{sec:conclusions}

In this work we have investigated PBH formation from large negative curvature perturbations, what we dubbed as Primordial Voids. Using numerical relativity simulations, we demonstrated that sufficiently deep PVs undergo a nonlinear rebound at their center that triggers black hole formation. We report the amplitude thresholds and PBH mass scaling for several equations of state. The thresholds lie in the type-II region fluctuations, while the PBH collapse formed in the vicinity after these thresholds correspond to type-A. For a comparative reference, we have also extended our investigation to the case of positive-amplitude fluctuations. 

This work represents one of the first explorations of the phenomenology associated with negative curvature perturbations generated during inflation. An important direction for future research is a more systematic study of the shape of the curvature profile. The profile can be modified by the width of the primordial power spectrum as well as by local non-Gaussianities, both of which influence the relative depth of the core and the height of the surrounding ridge. For sufficiently broad spectra, non-spherical features may also become relevant, and the statistics tend to favour sharper, shell–dominated configurations. A deeper understanding of how these shape variations affect the compaction function and the role of the three-Ricci curvature in the subsequent dynamics should ultimately improve analytical prescriptions for collapse thresholds for both positive and negative curvature profiles. We leave these extensions for future work.

A related open question concerns the possible existence of type-B PBHs sourced by PVs, corresponding to a ``separate universe'' configuration. If such objects can form, they would initially appear as trapped regions with an open-core geometry, and it would be important to clarify whether these evolve toward stable PBHs or instead disperse. A dedicated study would be required to address this scenario.

Furthermore, because underdensities or voids are necessarily present in any model that sufficiently enhances the scalar power spectrum, it is important to account for their associated phenomenology. Relevant examples include the generation of sound waves and the resulting gravitational wave signatures~\cite{Ning:2025ogq,Zeng:2025law}, potential effects during matter-dominated epochs such as those appearing in PBH reheating models~\cite{Allahverdi:2025btx}, or even their possible role in producing non-thermalized regions that act as catalysts for baryogenesis~\cite{Baumann:2007yr,Carr:2019hud,Garcia-Bellido:2019vlf,Despontin:2024jzp}. Understanding the signatures associated with relatively large positive and negative curvature fluctuations may help not only to constrain inflationary models at low frequencies, but also to elucidate potential mechanisms that address outstanding cosmological puzzles.

Finally, it would be valuable to extend the present analysis to scalar-field dominated epochs.
PBH formation shortly after inflation, whether during a phase of kination~\cite{Cheng:2025eas} or during (p)reheating~\cite{Khlopov:1985fch,Martin:2019nuw,Padilla:2025bkv,Milligan:2025zbu}, involves dynamics that may differ significantly from those of a perfect fluid. Incorporating the effects of coherent  scalar-field oscillations, scalar field gradients and nontrivial sound speeds would provide further insight on the viability of PBH formation (and other relics) during these early epochs.

\section*{Acknowledgments}
The authors thank Cristiano Germani, Shi Pi, Diego Cruces, S\'ebastien Clesse, Jaume Garriga, Shao-Jiang Wang, Xiang-Xi Zeng and Zhuan Ning for stimulating converstations. C. J. is supported by NSFC grants No. W2433007 and 12475066. 
Z.-Y. Y. is supported by an appointment to the Young Scientist Training (YST) program at the APCTP through the Science and Technology Promotion Fund and Lottery Fund of the Korean Government. This was also supported by the Korean Local Governments-Gyeongsangbuk-do Province and Pohang City.
Computational resources were borrowed from the SCNet High-Performance Computing Infrastructure, Sugon (Dawning Information Industry Co., Ltd.)

\appendix

\section{Details of the Numerical Relativity Simulations}
\label{appendix:NR_spherical}

\subsection{Evolution equations}

We solve the Einstein Field equations using the BSSN formalism, as described in Section \ref{sec:GRHydro}.

The evolution equations of the geometrical BSSN variables $\{\chi, a, b, K, A_a, \hat{\Delta}^r, \alpha, \beta^r\}$ are the following~\cite{Alcubierre:2011pkc}, 
\begin{eqnarray}
\partial_t \chi &=& \beta^r \partial_r \chi 
- \frac{1}{6} \alpha K \; , \\
\partial_t a &=& \beta^r \partial_r a + 2 a \partial_r \beta^r 
- 2 \alpha a A_a , \qquad \\
\partial_t b &=& \beta^r \partial_r b + 2 b \: \frac{\beta^r}{r} 
- 2 \alpha b A_b \; . \\
\partial_t K &=& \beta^r \partial_r K - \nabla^2 \alpha
+ \alpha \left( A_a^2 + 2 A_b^2 + \frac{1}{3} \: K^2\right) \nonumber  \label{eq:BSSN_K} \\
&+& 4 \pi \alpha \left( \rho_E + S_a + 2 S_b \right) , 
\\
\partial_t A_a &=& \beta^r \partial_r A_a - \left( \nabla^r \nabla_r \alpha
- \frac{1}{3} \nabla^2 \alpha \right)  \nonumber \\
&+& \alpha \left( ^{(3)}R^r_r - \frac{1}{3} ^{(3)}R \right) \nonumber \\
&+& \alpha K A_a - \frac{16}3 \pi \alpha \left( S_a - S_b \right) \; , 
\end{eqnarray}
\begin{eqnarray}
\partial_t \hD^r &=& \beta^r \partial_r \hD^r - \hD^r \partial_r \beta^r
+ \frac{1}{a} \partial^2_r \beta^r + \frac{2}{b} \:
\partial_r \left( \frac{\beta^r}{r} \right)  
\nonumber \\ &-& 
\frac{2}{a} \left( A_a \partial_r \alpha
+ \alpha \partial_r A_a \right)  \nonumber \\
&+& 2 \alpha \left( A_a \hD^r - \frac{2}{rb}
\left( A_a - A_b \right) \right) 
\nonumber \\ &+&   
\frac{2\alpha}{a} \left[ \partial_r A_a
- \frac{2}{3} \: \partial_r K  + 6 A_a \partial_r \chi 
\right. \nonumber \\
&+&  \left. \left( A_a - A_b \right) \left( \frac{2}{r}
+ \: \frac{\partial_r b}{b} \right) 
- 8 \pi S_r \right] \;  \label{eq:BSSN_Deltar} ,
\end{eqnarray}
where $\rho_E$ and $S_r$ are, respectively, the total energy density and the covariant component of the momentum density, as seen by an Eulerian observer, which are defined below. 
Note that the system of equations respects covariance, only that the terms of $\tilde\gamma^{rr} = a^{-1}$ and $\tilde\gamma^{\theta\theta} = \tilde\gamma^{\phi\phi} = 1/(br^2)$ have been explicitly written. 
The covariant derivatives with respect to the four-metric for the lapse function are computed as follows
\begin{eqnarray}
\nabla^2 \alpha &=&  \frac{1}{a e^{4 \chi}} \left[ \partial_r^2 \alpha
- \partial_r \alpha \left( \frac{\partial_r a}{2a}
- \frac{\partial_r b}{b}
- 2 \partial_r \chi - \frac{2}{r} \right) \right],  \nonumber
\\ 
\nabla^r \nabla_r \alpha &=& \frac{1}{a e^{4 \chi}} \left[ \partial_r^2 \alpha 
- \partial_r \alpha \left( \frac{\partial_r a}{2a}
+ 2 \partial_r \chi \right) \right] \; .
\end{eqnarray}

$^{(3)}R^r_r$ and $^{(3)}R$ are the diagonal-radial component of the Ricci tensor and the Ricci scalar, respectively. Assuming spherical symmetry, they read
\begin{align}
^{(3)}R^r_r = &- \frac{1}{a e^{4 \chi}} \biggl[ \frac{\partial^2_r a}{2a} 
- a \partial_r \hD^r - \frac{3}{4} \left( \frac{\partial_r a}{a} \right)^2
+ \frac{1}{2} \left( \frac{\partial_r b}{b} \right)^2 \nonumber \\
&- \frac{1}{2} \hD^r \partial_r a + \frac{\partial_r a}{rb} 
+ \frac{2}{r^2} \left( 1 - \frac{a}{b} \right) \left( 1 + \frac{r \partial_r b}{b} \right) \nonumber \\
&+ 4 \partial^2_r \chi - 2 \partial_r \chi \left( \frac{\partial_r a}{a} - \frac{\partial_r b}{b} - \frac{2}{r} \right) \biggr] , \label{eq:sphere-Rrr} 
\end{align}
\begin{align}
^{(3)}R = &- \frac{1}{a e^{4 \chi}} \biggl[ \frac{\partial^2_r a}{2a} + \frac{\partial^2_r b}{b} - a \partial_r \hD^r - \left( \frac{\partial_r a}{a} \right)^2  \\
&+ \frac{1}{2} \left( \frac{\partial_r b}{b} \right)^2 + \frac{2}{rb} \left( 3 - \frac{a}{b} \right) \partial_r b + \frac{4}{r^2} \left( 1 - \frac{a}{b} \right) \nonumber \\
&+ 8 \left( \partial^2_r \chi + ( \partial_r \chi )^2 \right) - 8 \partial_r \chi \left( \frac{\partial_r a}{2a} - \frac{\partial_r b}{b} - \frac{2}{r} \right) \biggr] . 
\nonumber
\end{align}

On the other hand, for the fluid dynamical system we use the framework described in Alcubierre's textbook \cite{10.1093/acprof:oso/9780199205677.001.0001}, which uses the following covariant  variables,
\begin{align}
 D &\equiv \rho_0 W  ~,  \label{eq-ap:conservedvars}
\\
 S_r &\equiv  (\rhofl + p) W^2 v_r  ~,   
\\
\mathcal{E} &\equiv  (\rhofl + p) W^2 -p - D    ~.    
\end{align}
Thus, is straightforward to identify the matter components in Eqs.(\ref{eq:BSSN_K}-\ref{eq:BSSN_Deltar}) as
\begin{align}
     \rho_E &= n^\mu n^\nu T_{\mu\nu} = D + \mathcal{E}  ~, \\
 S_i &= -\gamma^{\mu}_i n^\nu T_{\mu\nu} = S_r ~, \\
 S_{ij} &= \gamma^{\mu }_i \gamma^{\nu}_j T_{\mu\nu} ~,  \label{eq:Sij} 
\end{align}
where from Eq.(\ref{eq:Sij}) we obtain 
\begin{align}
    S_a &\equiv \gamma^{rr}S_{rr} = \frac 1 {a e^{4\chi}} (E+D+p) v_r^2 + p ~, \\ 
    S_b &\equiv \gamma^{\theta\theta}S_{\theta\theta} =  p ~.
\end{align}
The evolution equations for the fluid's conserved variables \{$E$, $D$, $S_r$\}, written in terms of the BSSN variables above, read
\begin{align}
\left( \partial_t -  \mathcal{L}_\beta \right) D  & =   - D_k (\alpha D v^k) + \alpha K D  ~,   \label{eq:evoD}
\\
 \left( \partial_t - \mathcal{L}_\beta \right) S^i  &=    - D_k \left[ \alpha  \left( S^i v^k  + \gamma^{ik} p \right) \right]   \nonumber \\ & \> - (\mathcal{E} + D) D^i\alpha  + \alpha K S^i  ~,     \label{eq:evoSi}
\\
\left( \partial_t - \mathcal{L}_\beta \right) \mathcal{E} & =
 (\mathcal{E} + D + p) (\alpha v^mv^n K_{mn} - v^m \partial_m \alpha) \nonumber \\
 & \>  - D_k \left[ \alpha v^k  \left( \mathcal{E} + p \right) \right]   + \alpha K (\mathcal{E} + p) ~,  \label{eq:evoE}
\end{align}
where $D_i$ denotes the covariant derivative with respect to the spatial metric and $\mathcal{L}_\beta$ the Lie derivative with respect to $\beta^i$. 
A generic equation of state often requires a root-finding technique to obtain the value of $p$ and $v^r$, prior to the recovery of the other variables, as they are needed to evolve Eqs.~\eqref{eq:evoD}--\eqref{eq:evoE}. However, for cases with a simple equation of state, the recovery of $p$ and $v^r$ can be done analytically by finding the physical root of a high-order polynomial. For a barotropic fluid with $p = \omega \rhofl $, this implies solving a second-order polynomial \cite{Staelens:2019sza}, which diminishes the computational burden of the simulations.  

\begin{figure*}[t!]
    \centering
    \includegraphics[width=0.95\linewidth]{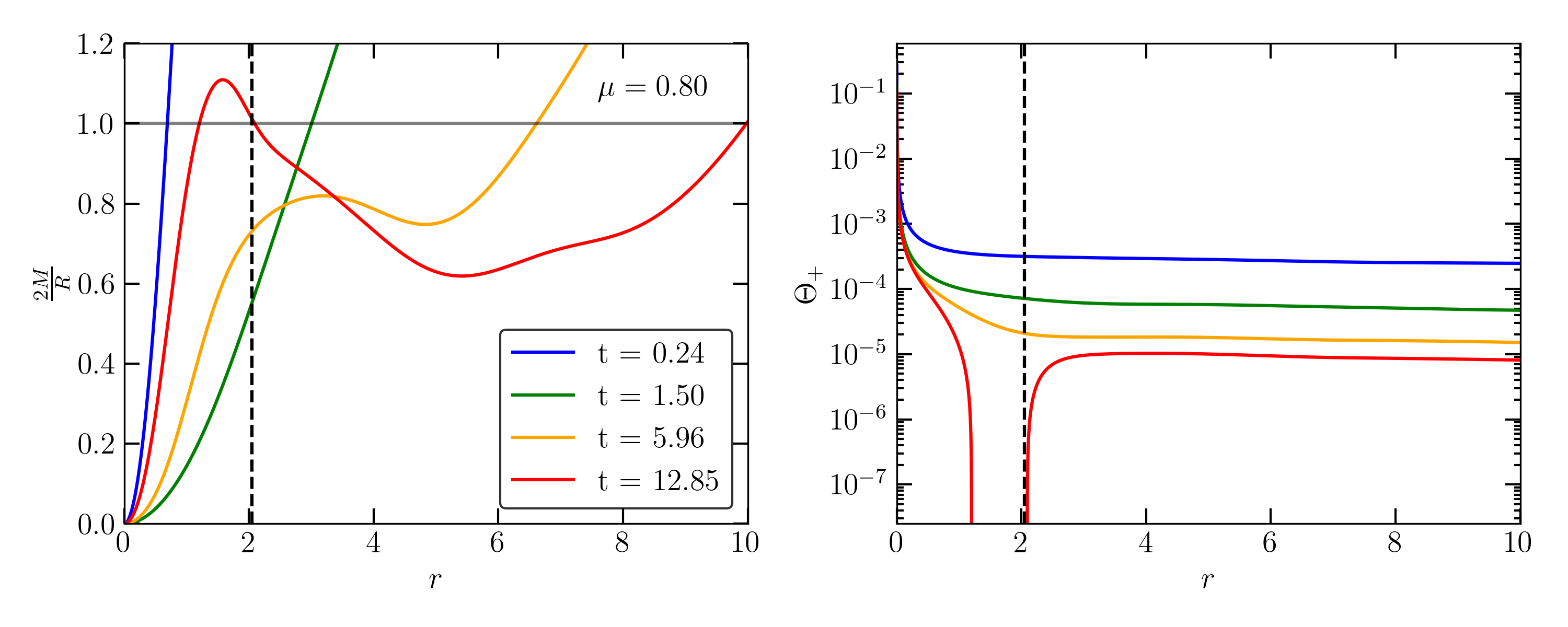}
    \includegraphics[width=0.95\linewidth]{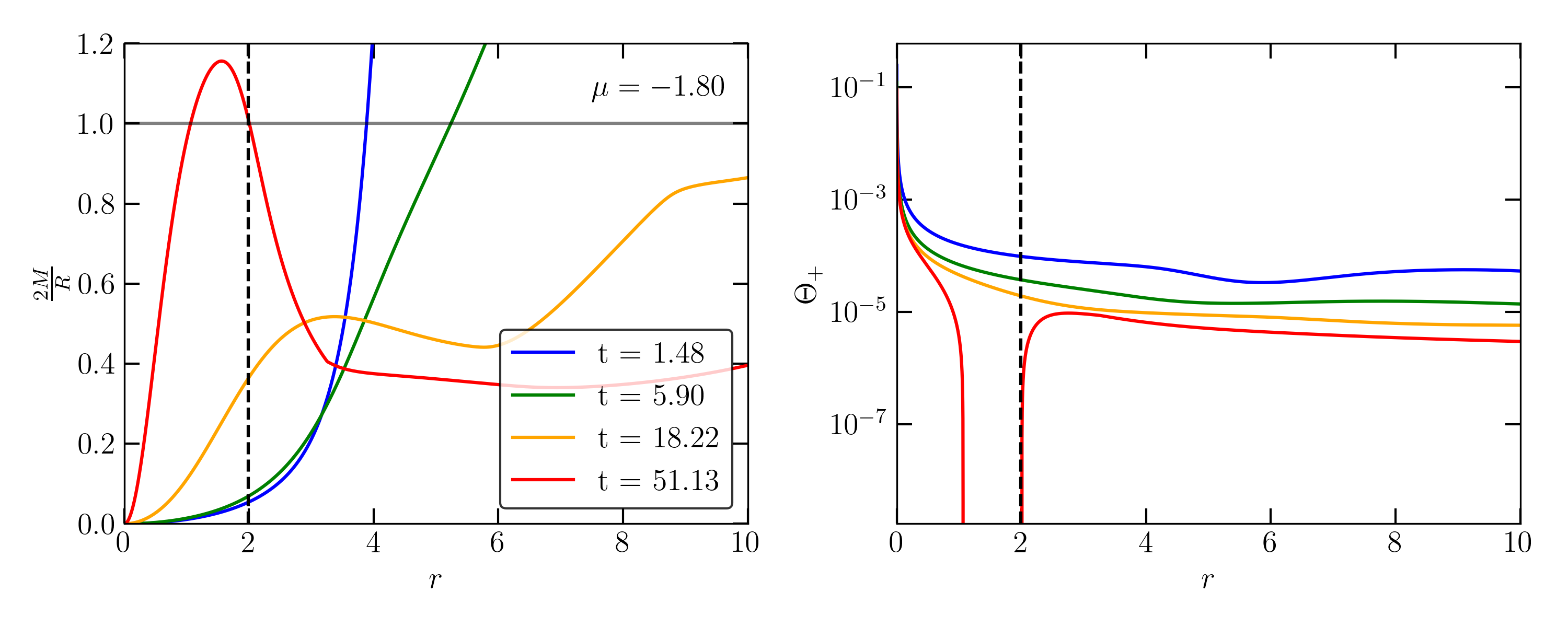}
    \caption{Formation and identification of the PBH's apparent horizon for $\mu= 0.8$ (top panels) and $\mu= -1.8$ (bottom panels)  by using the $2M_{\rm MS}/R = 1$ (left panels) and the $\Theta_{+}<0$ conditions (right panels). The vertical dashed black line marks the location of the AH, shortly after its formation (depicted by red solid curves). 
    }
    \label{fig:horizons}
    
    \vspace{0.5cm} 
    
    \centering
    \includegraphics[width=0.95\linewidth]{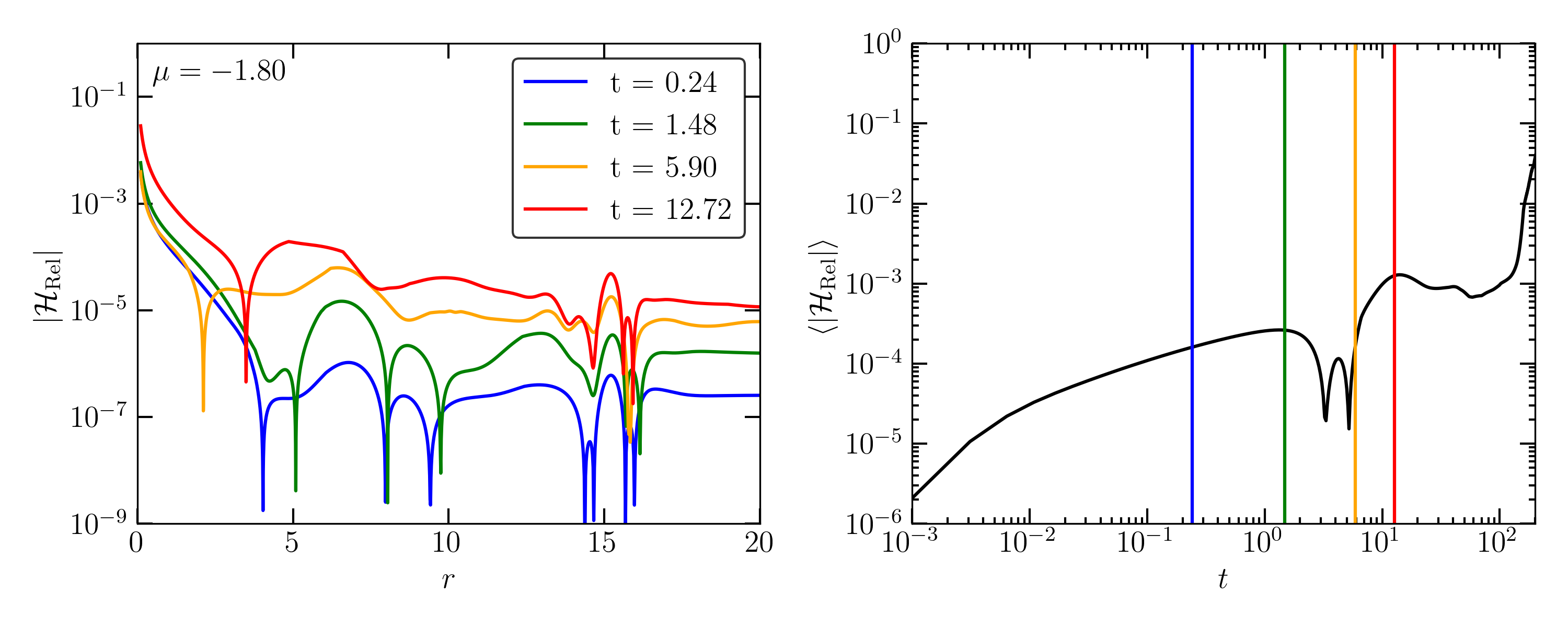}
    \caption{Violation of the relative Hamiltonian constraint. Left panel shows local values at specific time throughout the simulations. Right panel shows the grid-averaged values, where the vertical color lines denotes the timing of the configuration showed in the left panel.   
    }
    \label{fig:validation}
\end{figure*}

\subsection{Numerical Implementation}

We numerically integrate the system of equations using a fourth-order Runge-Kutta (RK4) method. The spatial derivatives are discretized on a uniform grid using a standard five-point stencil, which provides fourth-order accuracy for the interior points. As customary, to ensure numerical stability in the presence of high-frequency modes, we apply Kreiss-Oliger dissipation.

Our computational domain is defined in logarithmic radial coordinates. We employ a grid of between $N = 2000$ and $N = 6000$ cells. The outer boundary of the grid is set to a comoving distance of $r_{\text{max}} \sim 20\, k_*^{-1}$, which corresponds to a physical size of approximately $2000\, H^{-1}(t_0)$, ensuring it is sufficiently large to maintain a FLRW background at the outer boundary.

\subsection{Construction of initial data}

The initial data is computed by solving the Hamiltonian and Momentum constraints, 
\begin{align}
\mathcal{H} & = ~ ^{(3)}R + K^2-K_{ij}K^{ij}-16\pi \rho= 0\, , \label{eqn:HamSimp} \\
\mathcal{M}_i & = D^j (K_{ij} - \gamma_{ij} K) - 8\pi S_i =0\, , \label{eqn:MomSimp}
\end{align}
where $^{(3)}R$ is the 3-Ricci scalar. 

We assume a conformally flat initial hypersurface, corresponding to the BSSN variables $a=b=1$ and $A_a = A_b = \Delta^r = 0$. Under these assumptions, the previous equations are solve by using 
\begin{align}
    K &= - 3 H_{\rm bkg}(t_0) =   - \sqrt{24\pi \rho_{\rm bkg}(t_0)}  ~, \\
    \rho_E &= \rho_{\rm bkg}(t_0) + \delta\rho ~,\\
    \delta\rho &= ~ ^{(3)}R ~, \\
    S_i & = 0 ~,
\end{align}
where the background quantities are set to follow the FLRW relations of a perfect fluid
\begin{equation}
    H_{\rm bkg}(t) = \frac{2}{3(1+\omega)t}~,  \qquad \rho_{\rm bkg}(t) = \frac{3 H_{\rm bkg}^2(t)}{8\pi}~.
\end{equation}
We tested this choice of initial condition to another another popular choice based on gradient expansion \cite{Musco:2004ak,Harada:2015yda}, and we obtained identical results for the thresholds and PBH masses. However, because our method provides an exact solution to the constraint equations, rather than introducing deviations of order $\mathcal{O}(\epsilon^2)$ from the gradient expansion, the constraints are better satisfied throughout the numerical evolution.

\subsection{Mass and Apparent Horizons}

As described in section~\ref{subsec:trappedregions}, we can use both the Misner-Sharp relation (\ref{eq:AHcondition}) and the outgoing null expansion $\Theta_{+}$ to determine the formation of the PBH apparent horizon.  Using the BSSN system describe above, these quantities read
\begin{align}
    \Theta_{+} &=  \frac{e^{-2\chi}}{\sqrt{a}} \left( 4\, \partial_r \chi + \frac{2}{r} + \frac{ \partial_r b}{b} \right) + A_a - \frac{2}{3} K ~,  \\
     M_{\rm MS} &=  \frac{1}{2} R \left[ 1 + \left( \frac{\partial_t R - \beta^r \partial_r R}{\alpha} \right)^2   -  \frac{e^{-4\chi}} a  (\partial_r R)^2  \right]
    ~,
\end{align}
where 
\begin{align}
   \partial_t R &= e^{2\chi} r \sqrt{b} \left( 2 \,\partial_t \chi + \frac{1}{2} \frac{\partial_t b}{b} \right)  ~,\\
    \partial_r R & = e^{2\chi} \sqrt{b} \left(1 + 2 r\,\partial_r \chi 
     + \frac{r}{2} \frac{\partial_r b}{b} \right) ~.
\end{align}

Figure~\ref{fig:horizons} shows two examples of the AH formation and identification using these quantities for the cases $\mu = -1.80$, $\mu = -0.8$ with $\omega = 1/3$.

\subsection{Code validation}

As customary, we check for the validity of our simulations by checking the fulfillment of the constraint equations~(\ref{eqn:HamSimp}-\ref{eqn:MomSimp}). In particular, we use that relative violation of the Hamiltonian constraint be small, at least  $\left| \mathcal{H}_{\rm rel} < 10^{-2}\right|$, where 
\begin{equation}
    \mathcal{H}_{\rm rel}  \equiv ~\frac{ ^{(3)}R + K^2-K_{ij}K^{ij}-16\pi \rho } { |^{(3)}R| + |K^2| + |K_{ij}K^{ij}|  + | 16\pi \rho| }  ~.
\end{equation}
Larger violations deep inside the AH are ignored (near $r\sim 0$), as long stability outside the AH holds.  

Figure~\ref{fig:validation} \, shows an example of code validation for the case $\mu = -1.80$ with $\omega = 1/3$.

\bibliographystyle{apsrev4-1}
\bibliography{biblio.bib}
\end{document}